\documentclass{article}
\usepackage{spconf,amsmath,graphicx}
\usepackage{booktabs} 
\usepackage{amsfonts}

\title{Joint Music and Language Attention Models for Zero-shot Music Tagging}
\name{Xingjian Du$^{1}$,
      Zhesong Yu$^{1}$,
      Jiaju Lin$^{1}$,
       Bilei Zhu$^{1}$,
      Qiuqiang Kong$^{2}$
      }
\address{$^{1}$Bytedance   $^{2}$The Chinese University of Hongkong}

\begin{document}

\maketitle
\begin{abstract}

Music tagging is a task to predict the tags of music recordings. However, previous music tagging research primarily focuses on close-set music tagging tasks which can not be generalized to new tags. In this work, we propose a zero-shot music tagging system modeled by a joint music and language attention (JMLA) model to address the open-set music tagging problem. The JMLA model consists of an audio encoder modeled by a pretrained masked autoencoder and a decoder modeled by a Falcon7B. 
We introduce preceiver resampler to convert arbitrary length audio into fixed length embeddings. We introduce dense attention connections between encoder and decoder layers to improve the information flow between the encoder and decoder layers. We collect a large-scale music and description dataset from the internet. We propose to use ChatGPT to convert the raw descriptions into formalized and diverse descriptions to train the JMLA models. Our proposed JMLA system achieves a zero-shot audio tagging accuracy of $ 64.82\% $ on the GTZAN dataset, outperforming previous zero-shot systems and achieves comparable results to previous systems on the FMA and the MagnaTagATune datasets.

\end{abstract}
\section{Introduction}
\label{sec:intro}

Music tagging \cite{fu2010survey, choi2017convolutional} is a task to design a system that can automatically predict the tags of music recordings. Music tagging is an essential task in music information retrieval (MIR) and has attracted research interests in both academia and industry. Many existing music tagging systems focus on \textit{closed-set} tasks \cite{choi2017convolutional,choi2016automatic, won2020evaluation,won2021semi,li2023mert}, where the tags are predefined for each dataset. The closed-set music tagging tasks can be identifying the genres, moods, instruments, singers, and eras of music. However, those systems can not generalize to out-of-domain tags. In this work, we address \textit{open-set} music tagging tasks \cite{liu2022open} where the evaluation datasets have different tags from the training dataset.

Multi-modal large language modes (LLMs) have become a recent research hotspot \cite{li2023blip,dai2023instructblip,alayrac2022flamingo}, where connecting audio, image, and video with language models helps utilize their understanding and reasoning capabilities to better accomplish tasks such as classification and captioning. In this paper, we explore how LLMs can aid in improving the open-set music tagging task. One of the most important questions in multi-modal LLM research is how to allow language models to obtain multi-modal information, or in other words, how to connect multi-modal modules with LLMs. Current mainstream connection methods usually connect the last layer of the multi-modal encoder with the LLM decoder \cite{deshmukh2023pengi}. It is generally believed that the embedding of the last layer of the encoder contains more high-level semantic information, while the middle and low-level semantic information is ignored. However, music tagging tasks have a very wide distribution of label categories, where some labels require high-level semantics, such as genre and emotion, while others require middle and low-level semantics, such as instruments. Therefore, in the music tagging task, it may also be important to output the information of the middle layer of the audio encoder to the language model.

In this paper, we propose a new model called Joint Music and Language Attention Models (JMLAs). This model differs from previous multi-modal LLMs in that we designed a mechanism to perform cross-attention between multiple layers of the audio encoder and multiple layers of the LLM decoder, thereby enhancing the interaction of audio and text information at different semantic levels. This mechanism uses a module called the Perceiver Resampler, which introduces learnable parameters to the cross-attention, thereby bridging the gap between audio and language embeddings. Additionally, it can also reduce the dimensionality of the audio embedding to improve the computational efficiency of the JMLAs.

One challenge of training JMLAs is the lack of publicly available training data \cite{huang2022mulan}. To address this problem, we collect a large-scale music and description dataset from the internet. However, the raw descriptions may contain noisy and irrelevant information about the music. To address this problem, we propose to use ChatGPT \cite{gpt4} to process the raw descriptions into formalized and diverse captions. We show that JMLAs trained with the ChatGPT processed captions achieve better results than the JMLAs trained with raw descriptions.

This paper is organized as follows. Section \ref{section:JMLAs} introduces the joint music and language attention model. Section \ref{section:dataset} introduces the text generation with ChatGPT. Section \ref{section:experiments} shows experiments. Section \ref{section:conclusion} concludes this work.

\begin{figure*}
        \centering
        \vspace{2pt}
        \includegraphics[width=0.9\textwidth]{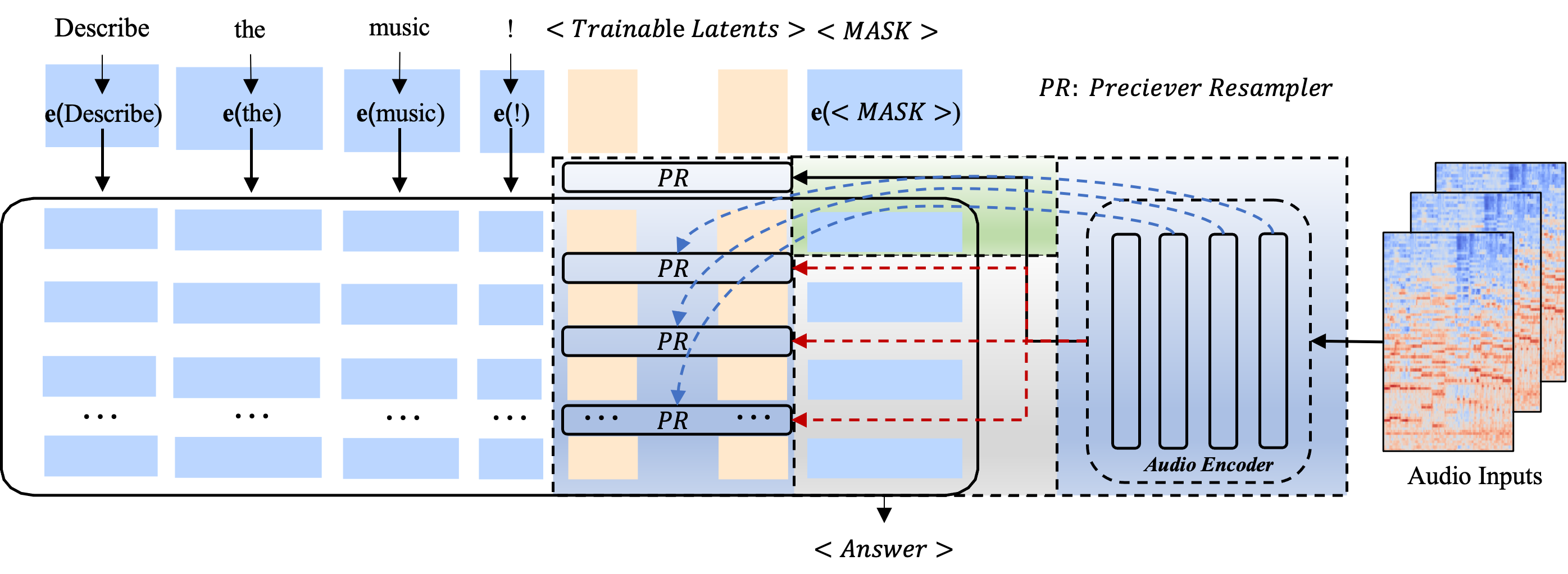}
        \vspace{-1em}
 \caption{The framework of the JMLA model, consists of an audio encoder, a language decoder, and perceiver resampler-based modules. The inputs to the model including audio and text. The output of the model are the decoded texts.  PR: perceiver resampler, uses the cross-attention mechanism to obtain a length-compressed audio representation for better efficiency and to decrease memory consumption while the LLM is processing the injected audio embedding. Red arrows: last-layer injection. Blue arrows: multiple layers injection.}
        \label{fig:framework}

\end{figure*}

\begin{figure}[t]
  \centering
  \vspace{-1em}
  \centerline{\includegraphics[width=\columnwidth]{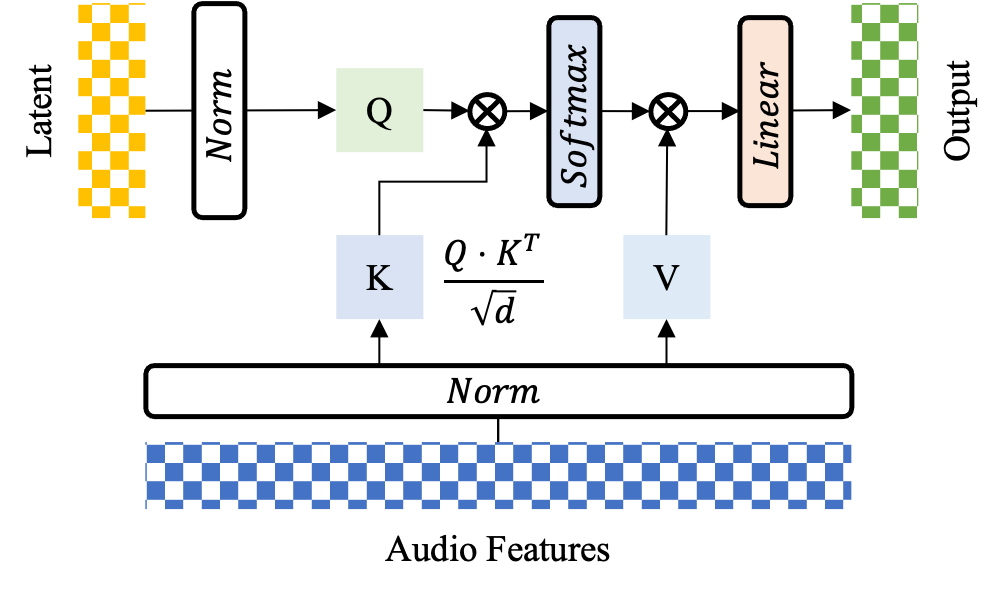}}
  \vspace{-1em}
  \caption{Perceiver resampler.}
  \vspace{-1em}
  \label{fig:perceiver_resampler}
\end{figure}

\section{Joint Music and Language Attention Models}\label{section:JMLAs}
Our proposed JMLA consists of an audio encoder, a language decoder, and an perceiver resampler-based multiple-layer attention module. Fig. \ref{fig:framework} shows the framework of the JMLA.

\subsection{Audio Encoder}

We first transform a time-domain waveform into log mel spectrogram $ X $ with a shape of $ T \times F $, where $ T $ is the frames number and $F$ is the frequency bins number. We train a masked autoencoder (MAE) \cite{niizumi2022masked} on the log mel spectrogram of a music recording and take the audio encoder of the MAE as the audio encoder. The log mel spectrogram are split into patches $ 16 \times 16 $ patches along the time and frequency axes. We denote the number of patches as $ P $. All patches are forwarded into fully connected layers.

We denote the MAE encoder output as $ \textbf{e} = f_{\text{enc}}(X) $ and the MAE decoder output $\hat{X} = f_{\text{dec}}(\textbf{e})$ where $ \hat{X} $ is the estimated spectrogram. During training, the MAE in applies a large position of masks that randomly remove 75\% of the patches in the input $ X $ \cite{niizumi2022masked}. In inference, we remain the MAE encoder layers and remove the MAE decoder layers to train JMLAs. The MAE encoder weights are frozen during the training of JMLAs.

\subsection{Perceiver Resampler}
The disadvantage of previous works \cite{deshmukh2023pengi} is that the computation cost increase quadratically with the length of audio. To address this problem, we propose to use a Perceiver resampler \cite{jaegle2021perceiver, alayrac2022flamingo} to convert arbitrary length audio embedding into a fixed length embedding. Fig. \ref{fig:perceiver_resampler} shows that the Perceiver uses a cross-attention module to project the audio embedding $ \textbf{e} $ with arbitrary length into a fixed length latent bottleneck $ \textbf{h} \in \mathbb{R}^{L \times D} $, where $ L $ is the sequence length of $ \textbf{h} $. More specifically, the key $ K $ and value $ V $ are projections of the audio embedding and query $ Q $ is a projection of a learned latent array with a length $ L \ll P $. The learned latent array contains learnable parameters during training.

\subsection{Multiple-layer Cross Attention}
In previous encoder-decoder architectures \cite{vaswani2017attention, ao2021speecht5}, the encoder output is directly input to the decoder. There is a lack of information flow between the encoder and decoder layers. To address this problem, we propose to introduce multiple layer connections between the intermediate layers of the encoder and deocder.

We introduce two types of multiple-layer cross attention in JMLA. The first type is to inject the \textit{last-layer output} of the audio embedding into individual perceiver resamplers and multiple decoder layers as shown in the red arrows in Fig. \ref{fig:framework}. The second type is to inject \textit{multiple-layer output} of the audio embedding into individual perceiver resamplers and multiple decoder layers as shown in the blue arrow in Fig. \ref{fig:framework}. By this means, there are more information flow between the audio encoder and language delcoder layers. 

To preserve the knowledge of the language models, we apply a prefix tuning \cite{li2021prefix} strategy that freeze the parameters of the language models part while only train the parameters of the percevier resamplers part. The yellow blocks of Fig. \ref{fig:framework} shows the trainable parameters. By this means, the JMLA model can maximize the utilization of the LLM.

\subsection{Language Decoder}

We input the audio representation $ \textbf{h} $ into the language decoder \cite{falcon40b} to autoregressively decode descriptions. 

The input to the language decoder is the concatenation of audio representation $ \textbf{h} $ and the texts of question $ \textbf{q} = \{q_{1}, ..., q_{M}\} $, where $ M $ is the number of words of the question. The target of the decoder is the texts of answer $ \textbf{d} = \{d_{1}, ..., d_{T}\} $, where $ N $ is the number of words of the description. 

The language encoder has a multiple layer causal Transformer architecture. We apply the pretrained Falcon7B \cite{falcon40b} as the language decoder. During the training of JMLA, the loss function can be written as:

\begin{equation} \label{eq:loss}
\mathcal{L} = -\sum_{t=1}^{T} \log(d_{t} | \textbf{e}, \textbf{q},d_{<t}).
\end{equation}
\noindent Equation (\ref{eq:loss}) shows that the JMLA autoregressively decode the texts of answer in a casual way.

\section{Dataset}\label{section:dataset}
\subsection{Music-description Dataset Collection}
Different from previous music tagging datasets \cite{tzanetakis_essl_cook_2001, defferrard2016fma, law2009evaluation} that only contain the close-set tags of music and due to the lack of publicly available music description datasets \cite{huang2022mulan}, we collect a large-scale dataset that contains natural language descriptions of music. The descriptions contains plentiful information of genre, speed, era, instrument, emotion, key of the music. We crawl music data from the inhouse data store. We collected 1.5 million audios and description pairs.

\subsection{Dataset processing with ChatGPT}

There are two main problems of the raw descriptions of the downloaded dataset. First, although the descriptions are ablundant in quantity, the descriptions can be noisy and irrelevant to the music recordings. For example, some descriptions may focus on the historical background of the music while others concentrate on the instrument analysis. Such dissimilarity in expression will result in difficulty in training. To address this problem, we format the descriptions into question-answer formats by using ChatGPT \cite{gpt4}. We design prompts and ask GPT-3.5-turbo to generate question-answer pairs based on given descriptions to increase the diversity.

\section{Experiments}\label{section:experiments}

\subsection{Datasets}

We evaluate our proposed music language model on diverse audio tagging datasets to compare our music language model with other systems, including the GTZAN \cite{tzanetakis_essl_cook_2001}, Free Music Archive (FMA) \cite{defferrard2016fma}, and the MagnaTagATune datasets \cite{law2009evaluation}. The GTZAN dataset consists of 1,000 30-second audio clips with 10 genres. The FMA dataset FMA dataset small set contains 8,000 audio clips with 8 genres. The MagnaTagATune dataset contains 25,863 30-second clips with 188 genres. Due to the large number of labels in this dataset, the top 50 most commonly used labels are usually employed. We adopt the classification accuracy as the evaluation metric.

\subsection{Zero-shot Music Tagging Results}

\begin{table}
  \caption{Zero-shot tagging results with different systems.}
  \vspace{2pt}
  \label{table:results_systems}
  \resizebox{\columnwidth}{!}{%
  \centering
  \vspace{-2em}
  \begin{tabular}{lcccc}
    \toprule
    & \multicolumn{1}{c}{\textbf{GTZAN}} & \multicolumn{1}{c}{\textbf{FMA}} & \multicolumn{2}{c}{\textbf{MagnaTagATune}} \\
    \cmidrule(lr){2-2} \cmidrule(lr){3-3} \cmidrule(lr){4-5}
    & Acc & Acc & PR & AUC \\
    \midrule
    AudioFlamingo \cite{alayrac2022flamingo} $ \dag $ & 38.62 & 39.62 & 17.43 & 61.30 \\
 Pengi \cite{deshmukh2023pengi} & 32.25 & - & - & - \\
 CLAP HTS-AT \cite{chen2022hts} $ \dag $ & 57.24 & 45.38 & 31.73 & 79.60 \\
 CLAP MAE (MuLan) \cite{huang2022mulan} $ \dag $ & 60.04 & 33.00 & 10.63 & 64.41 \\
 \midrule
 JMLA & 58.28 & 41.88 & 18.64 & 70.66 \\
 JMLA-DenseDec & 60.34 & 42.00 & 21.06 & 74.42 \\
 JMLA-DenseEncDec & 64.82 & 41.00 & 20.78 & 73.51 \\
\bottomrule
\end{tabular}}
\footnotesize The $\dag$ symbol indicates our reproduced system.
\end{table}

\begin{table*}
  \caption{Zero-shot tagging results with different promots.}
  \vspace{4pt}
  \label{table:results_promot}
  \resizebox{\textwidth}{!}{%
  \centering
  \vspace{-2em}
  \begin{tabular}{llllcccc}
    \toprule
    & & & & \multicolumn{1}{c}{\textbf{GTZAN}} & \multicolumn{1}{c}{\textbf{FMA}} \\
    \cmidrule(lr){5-5} \cmidrule(lr){6-6} 
    Input & Input Example & Output & Postprocess & Acc & Acc \\
    \midrule
    Prompt & What is the genre of the music? & Sentence & Similarity & 35.80 & 22.25\\
    Promot + TagsList & What is the genre of the music? The genres include pop, blues, ... & Sentence & Similarity & 40.80 & 35.00\\
    Promot + TagsList & What is the genre of the music? Answer one word from pop, blues, ... & OneHot & N/A & 56.55 & 39.75\\
    Promot + All Candidates & What is the genre of the music? The answer is $\{\text{pop}~|~\text{blues}~|~\text{...}\}$ . & N/A & Log-likelihood  & 64.82 & 40.50\\
\bottomrule
\end{tabular}}
\end{table*}

\begin{table}
  \caption{Zero-shot tagging results with different training data.}
  \vspace{2pt}
  \label{table:results_training_data}
  \resizebox{\columnwidth}{!}{%
  \centering
  \vspace{-2em}
  \begin{tabular}{lcccc}
    \toprule
    & \multicolumn{1}{c}{\textbf{GTZAN}} & \multicolumn{1}{c}{\textbf{FMA}} & \multicolumn{2}{c}{\textbf{MagnaTagATune}} \\
    \cmidrule(lr){2-2} \cmidrule(lr){3-3} \cmidrule(lr){4-5}
    & Acc & Acc & PR & AUC \\
    \midrule
    RawCaption & 50.68 & 28.87 & 25.58 & 77.37\\
    GPT-QA & 60.34& 42.00 & 21.06 & 74.42\\
    RawCaption + GPT-QA & 53.79 & 41.87 & 23.90 & 77.12\\
    GPT-QA Finetune & 62.40 & 43.60 & 23.63 & 74.30\\
\bottomrule
\end{tabular}}
\end{table}

Table \ref{table:results_systems} shows the music tagging results of different systems. We compare our system with the Audio Flamingo \cite{alayrac2022flamingo}, Pengi \cite{deshmukh2023pengi}, CLAP HTS-AT \cite{chen2022hts}, and the CLAP MAE \cite{huang2022mulan} systems using the same data as our system. All systems are evaluated in a zero-shot way. That is, only the evaluation subset of the datasets are used, without trained or finetuned using the training subset of the datasets.

\noindent (1) \textbf{AudioFlamingo} \cite{alayrac2022flamingo} is a multimodal architecture, with a cross-attention module and resampling blocks. Flamingo is trained on large-scale multimodal web corpora containing arbitrarily interleaved text and images.

\noindent (2) \textbf{Pengi} \cite{deshmukh2023pengi} is a novel audio language model that leverages transfer learning by framing all audio tasks as text-generation task. In Pengi, audio embedding sequences are combined as a prefix to prompt a pre-trained frozen language model. We use the officially released model weights for evaluation.

\noindent (3) \textbf{CLAP-HTS-AT}\cite{chen2022hts} is a language-audio model that aligns language and audio representation in the same space by contrastive learning. The model is trained on 633,526 audio-text pairs from different data sources. We use the HTS-AT version, where the audio encoder is realized by HTS-AT.

\noindent (4) \textbf{CLAP-MAE} is our re-implementation of CLAP. Regarding the limited representation ability of HTS-AT, we substitute Audio MAE for HTS-AT as the audio encoder. For fair comparison, we re-train the system with our music-description dataset. 

In order to evaluate the correctness of the downstream classification task, we employ a vocabulary ranking method following previous work \cite{deshmukh2023pengi}. We concatenate $N$ possible candidates after the question and constructing $N$ complete sentences. The highest log-likelihood score of each candidates is select as the final predicted result.

Table \ref{table:results_systems} shows that the AudioFlagmingo achieves an accuracy of 38.62\% on the GTZAN dataset. The Pengi system achieves a lower accuracy of 32.25\%. The CLAP HTS-AT or the CLAP-MAE system improve the accuracy to 57.24\% and 60.04\%, respectively. Our proposed JMLA system without multiple-layer attention achieves an accuracy of 58.28\%. Our proposed JMLA-DenseDec refers to the system that the output of the encoder is input to multiple decoder layers improves the accuracy to 60.34\%. Our proposed JMLA-DenseEncDec refers to the system that the mutliple layer output of the encoder are input to multiple decoder layers improves tehe accuracy to 64.82\%. On the FMA dataset, the JMLA system achieves an accuracy of 41.00\%, outperforming the AudioFlamingo and CLAP MAE system, while slightly underperforms the CLAP HTS-AT system. The MagnaTagATune dataset is a more challenging task than the GTZAN and the FMA datasets due to there are 50 genres. The JMLA system achieves a PR of 20.78\%, outperforming the AudioFlamingo and the CLAP MAE system.

\subsection{Different Promots Results}

Table \ref{table:results_promot} shows the JMLA results with different promots. The promot consists of a question and the token of the audio. Table \ref{table:results_promot} shows that different types of promots will lead to different results. By limiting the predicted genres to be one of the genres in the dataset, the accuracy improves from 35.80\% to 40.80\% and 56.55\%, respectively. In addition, we show that by taking the predicting with log-likelihood the accuracy improves to 64.82\%. This results indicate designing promots are important to the zero-shot music tagging. 
 
\subsection{Different Training Data Results}
Table \ref{table:results_training_data} shows the results of JMLA trained with four types of data: 1) \textbf{Raw caption} data crawled from the internet; 2) \textbf{GPT-QA} data created by using ChatGPT to process and clean the raw caption data; 3) \textbf{RawCaption + GPT-QA} data that concatenates the RawCaption and the GPT-QA data; 4) \textbf{GPT-QA Finetune} data that use GPT-QA data to finetune the system pretrained with RawCaption + GPT-QA data.

Table \ref{table:results_training_data} shows that the JMLA system trained with the RawCaption data achieves the lowest accuracy of 50.68\% on the GTZAN dataset. When training on the GPT-QA dataset, the accuracy significantly improves the accuracy to 60.34\%. This result shows the effectiveness of the GPT-QA data to train the JMLA system. Table \ref{table:results_training_data} shows that when training on the concatenation of RawCaption + GPT-QA data, the accuracy decreases to 53.79\%. This result shows that the RawCaption data has negative effect to train the JMLA system. Furthermore, we show that after finetuning the RawCaption + GPT-QA system with the GPT-QA data, the accuracy further improves to 62.40\%.

\vspace{-1em}
\section{Conclusion}\label{section:conclusion}
In this work, we propose a joint music language attention model (JMLA) model consists of an audio encoder modeled by audio masked auto-encoder (MAE), a perceiver and mutliple-layer attention module, and a language decoder. Our proposed JMLA can address the open vocabulary audio tagging tasks. We collect a large-scale music language dataset from the website and process the data with a ChatGPT to process the dataset. We show that the JMLA outperforms and achieves comparable results to previous zero-shot audio tagging system on the GTZAN, FMA, and MagnaTagATune datasets, respectively. In future, we will investigate more music question answer problems with JMLA.

\small
\bibliographystyle{IEEEbib}
\bibliography{strings,refs}

\end{document}